# Synthesis of self-assembled iron silicon oxide nanowires onto single-crystalline Si(100)†

Guotao Lu[*]

*Department of Chemistry and Graduate Center for Materials Research, University of Missouri-Rolla, Rolla, Missouri 65409-110, USA.*

**Abstract:**

Iron silicon oxide nanowire has been synthesized by putting single-crystalline Si(100) wafer into ammonium iron sulfate $((NH_4)_2Fe(SO_4)_2 \cdot 9H_2O)$, hydrogen peroxide $(H_2O_2)$ and triethylamine (TEA) solution heated to 70-100 C° in the air. The prepared iron silicon oxide nanowires with diameters of ~10nm (onto no-etched Si(100) wafer) and ~75nm (onto etched Si(100) wafer) and their self-assembling were characterized. The possible mechanism that accounted for the formation of iron silicon oxide nanowire is suggested based on experimental evidence.

**Introduction**

Recently, nanowires have received a lot of attention because of their nano-scale one-dimensional structure and unique mechanical, electrical, magnetic properties, as well as their potential applications for various mesoscopic electronic and optical devices.[1,2] Si-containing nanowires, such as amorphous and crystalline silicon oxide, silicon and doped silicon nanowires, are of great interest based on their vital roles in nanoscale electronics, sensors, optoelectronics and photoluminescent materials.[3-6]

There are several methods to prepare Si-containing nanowires. Among them, the traditional one is chemical vapor deposition;[7] the others include the vapor-liquid-solution method,[8] the laser vaporization method,[9] and the solution growth.[10] Holmes *et al.* have made defect-free silicon (Si) nanowires with a supercritical fluid solution-phase approach.[10] Even though, there are few papers

---

reported the common solution-phase synthesis of Si-containing nanowires. In this paper, I report a new hydrothermal solution-phase preparation of iron silicon oxide nanowires and present a plausible mechanism for the formation of these nanowires.

Iron silicon oxide ($Fe_7SiO_{10}$) was first isolated in 1969 by Smuts *et al.* from a furnace refractory brick.[11] The unit cell was monoclinic, with a = 2.14 nm, b = 0.306 nm, c = 0.588 nm, β=98°.[11] The structure consisted of a stacking of FeO and $Fe_3SiO_6$ blocks. Later, Tilley *et al.* refined its composition with $Fe_7(Si_{0.94}Fe_{0.06})O_{10}$ and found a paramagnetism to a weak ferromagnetism transition at ~250 K which showed that $Fe_7SiO_{10}$ is a ferromagnetic semiconductor below ~250 K.[12] Bhagwat *et al.* have found the presence of $Fe_7SiO_{10}$ at the interface of Fe and $SiO_2$ interface.[13] So far, $Fe_7SiO_{10}$ has potential applications as ferromagnetic semiconductors[14] and magnetic-optical films.[15]

**Experimental**

The iron silicon oxide nanowires were prepared onto two types of substrates: etched and non-ethched p-type Si(001) wafers (Virginia Semiconductor). The etched Si(100) wafers were prepared by putting it into 30% HF solution for 24 hours and then rinsing with ethanol and deionized (DI) water. The typical procedure is as follow: 20.5 g $(NH_4)_2Fe(SO_4)_2·9H_2O$ (0.05M) and 5.05g Triethylamine was dissolved into 100ml DI water, and heated the solution to 70 ºC, then the Si(100) wafer held by a PTFE clamper was put into the solution. 40ml 30% $H_2O_2$ was slowly drop-added into the solution within 1hr. For the etched Si(100) substrate, the reaction temperature was 96 ºC to promote the formation of $SiO_2$ layer and iron silicon oxide nanowires. The reaction was continued for another hour and some brown precipitations occured in the solution. The resulting film on the Si(100) surface was grey-brown and covered the whole wafer. The film was rinsed with ethanol and DI water several times to remove the residual TEA and dried in an oven at 60ºC. The p-type Si(001) wafers were supplied by Virginia Semiconductor, and doped with boron to a resistivity of 7.5 Ωcm, and a hole density of about $2 \times 10^{15}$ $cm^{-3}$. The sample was examined by X-ray diffraction measurements with a high-resolution Philips X'Pert MRD diffractometer. For the Bragg-Brentano scan, the primary optics module was a combination Gobel mirror and a two-crystal Ge(220) two-bounce hybrid monochromator which produces pure $CuK\alpha_1$ radiation (λ = 0.1540562 nm) with a divergence of 25 arc sec. A 0.18º parallel plate collimator served as the secondary optics. FT-IR spectra were recorded on a Magna-IR™ Spectrometer 750 (Nicolet

Analyticl) with 512 scans under nitrogen at a resolution of 2 cm$^{-1}$. The microstructures of the films were characterized by SEM micrographs with a Hitachi model S4700 cold field-emission scanning electron microscope and energy dispersive X-ray (EDX) analysis available on this SEM.

**Results and Discussions**

Fig. 1 shows the FE-SEM photographs of iron silicon oxide nanowires onto a non-etched Si(100) wafer. It can be seen from the figure that iron silicon oxide nanowires with diameters of ~10nm were formed and they are self-assembled into a tree-like network. More FE-SEM pictures were given in Electronic Supplementary Information† (Fig. S1). EDX studies were performed directly on the film on the Si(100) wafer. And the ratio of Fe: Si: O are 1: 3.2: 3.4(ESI† Tab. S1). The reason for differences from the stoichiometric ratio is that the substrate was a Si wafer which will enhance the amount of Si compared with $Fe_7SiO_{10}$ formula. The uncovered Si wafer surface and the in-situ produced $O_2$ will also react with Si substrate to produce $SiO_2$. The reactions accounting for the formation of $Fe_7SiO_{10}$ and $SiO_2$ are shown as follows:

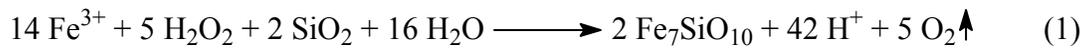

$$14\ Fe^{3+} + 5\ H_2O_2 + 2\ SiO_2 + 16\ H_2O \longrightarrow 2\ Fe_7SiO_{10} + 42\ H^+ + 5\ O_2\uparrow \quad (1)$$

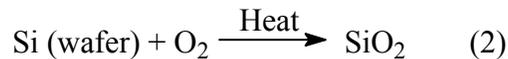

$$Si\ (wafer) + O_2 \xrightarrow{Heat} SiO_2 \quad (2)$$

Then the ratio of Fe: Si: O can be explained by the existence of $Fe_7SiO_{10}$, Si and $SiO_2$. The formation of $Fe_7SiO_{10}$ was also examined by X-ray diffraction. Fig. 2 shows the XRD pattern of iron silicon oxide nanowires onto non-etched Si(100) wafer. Except for the peak from Si(100) substrate with $2\theta = 33.0°$( Si(200)) and $2\theta = 70.3°$(Si(400)), the strongest peak from the film is at $2\theta = 42.6°$, and all the other peaks are not strong enough to be used to determine the compound. Through carefully checking the data with JCPDS, we think that the iron slilicon oxide is $Fe_7SiO_{10}$ (JCPDS #78-1653). The peak at 42.6° was assigned to $Fe_7SiO_{10}$ ($\underline{1}$12). This result is consistant with that from EDX.

FT-IR spectra of iron silicon oxide nanowires, deposited onto a non-etched Si(100) wafer, are shown in Fig. 3. It can be seen in Fig. 3 that the band at 1080 cm$^{-1}$ associated with asymmetrical stretching Si-O-Si mode ($AS_1$) for $SiO_2$ did not change before and after the deposition of $Fe_7SiO_{10}$ nanowires onto the non-etched Si(100) wafer.[16-17] The intensity of the Si-OH stretching vibration at 950 cm$^{-1}$ was lowered a lot after $Fe_7SiO_{10}$ nanowires were deposited onto non-etched Si(100) wafer.[21-22] This means that because of the reaction, the amount of $SiO_2$ decreased, which was also can be observed from SEM image of the $Fe_7SiO_{10}$ nanowires film(ESI† Fig. S1(a)). In Fig. S1(a), it can be seen that in the middle, there still exists a layer of $SiO_2$ with thickness of about 50nm, whereas at most surface of the Si wafer, the $SiO_2$ layer had already disappeared through the reaction (1). The Si(100) wafer used in this experiment was non-etched and has a native amorphous SiO2 layer at about 50 nm which was consistent with our SEM result.[18] The band at 800 $cm^{-1}$ associated with symmetrical stretching Si-O-Si mode($S_1$) for $SiO_2$ did not change before and after the deposition. The formation mechanism for this tree-like self-assemble nanowire film will be suggested at the end of the paper after the discussion of the preparation of $Fe_7SiO_{10}$ nanowires deposited onto an etched Si(100) wafer.

The purpose of the etching of the Si(100) wafer with 30% HF solution is to remove the native amorphous $SiO_2$ layer on the Si wafer surface and to change the Si-OH to stable Si-H group. Fig. 6 (a) shows the FT-IR spectroscopy of the etched Si(100) wafer. It is obvious that the $SiO_2$ amount drops dramatically and no vibration at 950 $cm^{-1}$ is observed from Si-OH. Many researchers have observed that after HF etching, the $SiO_2$ layer on the Si wafer surface decreased to about ~5 nm.[19] This very thin $SiO_2$ layer will affect the resulting $Fe_7SiO_{10}$ nanowires' morphology in significant way.

Fig. 4 shows the FE-SEM photographs of the as-prepared $Fe_7SiO_{10}$ nanowires film. It can be seen from the figure that nanowires with diameterd of ~ 60nm were formed and were connected to form a network. More SEM images are available from ESI† Fig. S3. EDX data was also shown in ESI† Fig. S4 and Tab. 2. The ratio of Fe: Si: O in this film is 1:39.4:3.4, which suggested a very small amount of $Fe_7SiO_{10}$ nanowires in the film. Most of the Si came from the Si(100) wafer. This situation is different from that of the non-etched Si(100) wafer in which half of Si came from $SiO_2$. The thin layer of $Fe_7SiO_{10}$ nanowires film was also verified by Fig. 7. This will be discussed later. The X-ray diffraction is shown in Fig. 5. The XRD pattern is almost the same as that of the $Fe_7SiO_{10}$ nanowires film deposited onto non-etched Si(100) wafer. As presented before, this pattern was assigned to $Fe_7SiO_{10}$ ($\underline{1}$12).

FT-IR spectra of $Fe_7SiO_{10}$ nanowires film (b) with the etched Si(100) wafer (a) are presented in Fig. 6. As discussed above, the etched Si wafer has a low amount of $SiO_2$ and Si-OH was effectively absent, which means the Si-OH group was not present in the $SiO_2$ layer surface, while the Si-H group was. After deposition of the $Fe_7SiO_{10}$ nanowires, the intensity at 1080 $cm^{-1}$ associated with the asymmetrical stretching of the Si-O-Si mode ($AS_1$) for $SiO_2$ increased dramatically, and there was the same trend in the 800 cm$^{-1}$ associated with the symmetrical stretching Si-O-Si mode for $SiO_2$. This suggests that a new $SiO_2$ layer is formed during the process via reaction (2). The band at 950 cm$^{-1}$ associated with Si-OH stretching vibration appeared in the $Fe_7SiO_{10}$ nanowires film, which also suggested the formation of a new $SiO_2$ layer. The excess amount of $H_2O_2$ in the solution producef excess amount of $O_2$ either through oxidation by Fe(III) ions or its own disproportionation which will easily react with the Si-H or Si wafer to generate the $SiO_2$ layer.

To understand the mechanism of the growth of $Fe_7SiO_{10}$ nanowires, a high magnification FE-SEM image was analyzed in Fig. 7. It can be observed from that figure that a $SiO_2$ layer with a thickness about ~20 nm deposited onto the Si(100) substrate, and within and above the $SiO_2$ layer, a $Fe_7SiO_{10}$ nanowire network formed. The diameter of the $Fe_7SiO_{10}$ nanowires was about 60nm, which is much higher than that of $Fe_7SiO_{10}$ nanowires deposited onto non-etched Si(100) wafer. The reason for this is that in the etched Si(100) wafer, the $SiO_2$ layer is very thin, higher temperature (96ºC) was used to perform the reaction than was used for the non-etched Si(100) wafer (70ºC). It is understandable that in the previous conditions, much thicker $Fe_7SiO_{10}$ nanowires were obtained than the latter.

Based on all the experiment evidence, we propose a growth model to account for the formation of iron silicon oxide nanowires, as shown in Fig. 8. As seen in the model for non-etched Si wafers, Fe(III) and $H_2O_2$ reacted with the $SiO_2$ layer to form nanowires within the layer by using triethylamine as a template, and most of the $SiO_2$ in the layer gradually reacted with the solution to form brown $Fe_2Si_2O_6$ precipitation. At the end, only the tree-like $Fe_7SiO_{10}$ nanowires cluster onto the Si(100) wafer surface remained, just as shown in the Fig. 1. For the etched Si wafer, the $SiO_2$ layer was first formed with a thickness of about ~20nm, which takes much time. Then the next step is the same as with the non-etched Si(100) wafer situation. But in this situation the $Fe_7SiO_{10}$ nanowires just formed the network and there is no enough time for the new-formed $SiO_2$ layer to react with the Fe(II) produced by reduction reaction of the Fe(III) to form $Fe_2Si_2O_6$, then at the same reaction time, in this situation, ended with $Fe_7SiO_{10}$ nanowires within and above the $SiO_2$ layer, as shown in Fig. 7. If the reaction lasted for long time for the etched Si wafer, it will be supposed to get the same result with the non-etched Si wafer. The formation of the highly-ordered $Fe_7SiO_{10}$ nanowires structures could be driven by triethylamine, which is well known as a self-assemble template.[20]

In conclusion, $Fe_7SiO_{10}$ nanowires self-assemble films have been deposited onto etched and non-etched Si(100) wafers. The $SiO_2$ layer is critical to form $Fe_7SiO_{10}$ nanowires, and triethylamine also plays a vital role for both the formation of the $Fe_7SiO_{10}$ nanowires and their self-assembling. The magnetic properties and its potential applications in ferromagnetic semiconductors and other areas will be examined in our future research.

**References**

1 H. Kind, H. Q. Yan, B. Messer, M. Law and P. D. Yang, *Adv. Mater.*, 2002, **14**, 158.

2 J. F. Wang, M. S. Gudiksen, X. F. Duan, Y. Cui and C. M. Lieber, *Science*, 2001, **293**, 1445.

3 J. Hu, T. W. Odom, and C. M. Lieber, *Acc. Chem. Res.*, 1999, **32**, 435.

4 J. Zhou, N. S. Xu, S. Z. Deng, J. Chen, J. C. She and Z. L. Wang, *Adv. Mater.*, 2003, **15**, 1835.

5 J. T. Hu, O. Y. Min, P. D. Yang and C. M. Lieber, *Nature*, 1999, **399**, 48.

6 M. S. Gudiksen, L. J. Lauhon, J. Wang, D. C. Smith and C. M. Lieber, *Nature*, 2002, **415**, 617.

7 X. Q. Yan, D. F. Liu, L. J. Ci, J. X. Wang, Z. P. Zhou, H. J. Yuan, L. Song, Y. Gao, L. F. Liu, W. Y. Zhou, G. Wang and S. S. Xie, *J. Cryst. Growth*, 2003, **257**, 69.

8 M. K. Sunkara, S. Sharma, R. Miranda, G. Lian and E. C. Dickey, *Appl. Phys. Lett.*, 2001, **79**, 1546.

9 A. M. Morales and C. M. Lieber, *Science*, 1998, **279**, 208.

10 J. D. Holmes, K. P. Johnston, R. C. Doty and B. A. Korgel, *Science*, 2000, **287**, 1471

11 J. Smuts, J. Steyn, and J. Boeyens, *Acta Crystallogr.*, 1969, **B 25**, 1251.

12 A. Modaressi, B. Malaman, C. Gleitzer and R. J. D. Tilly, *J. Solid. State Chem.*, 1985, **60**, 107.

13 S. Bhagwat, S. N. Yedave, D. M. Phase, S. M. Chaudhari, S. M. Kanetkar and S.

B.Ogale, *Phy. Rev. B*, 1989, **40**, 700.

14 S. J. Pearton, C. R. Abernathy, M. E. Overberg, G. T. Thaler, D. P. Norton, N. Theodoropoulou, A. F. Hebard, Y. D. Park, F. Ren, J. Kim, L. A. Boatner, *J. Appl. Phys.* 2003, **93**, 1.

15 Y. Togami, T. Morishita and K. Tsuchima, *Jpn. J. Appl. Phys.*, 1987, **26**, L258.

16 D.V. Tsu and G.Lucovsky, and B. N. Davidson. *Phys.Rev. B*, 1989, **40**, 1795.

17 C.T. Kirk, *Phys.Rev. B*, 1988, **38**, 1255.

18 B. Okolo, P. Lamparter, U. Weizel and E. J. Mittemeijer, *J. Appl. Phys.*, 2004, **95**, 466.

19 J. A. Switzer, R. Liu, E. W. Bohannan and F. Ernest, *J. Phys. Chem. B.*, 2002, **106,** 12369.

20 R. S. Sapieszko and E. Matijevic, *J. Colloid Interface Sci.*, 1980, *74*, 405.

21. G. Lu and Y. Huang, *J. Mater. Sci. 2002,* **37**(11), 2305-2309.

22. G. Lu, Y. Huang, Y. Yan, T, Zhao and Y. Yu, *J. Polym. Sci. Part A:Polym. Chem.,* 2003, **41**(16), 2599-2606.

**Figure captions**

Figure 1. FESEM of iron silicon oxide nanowires deposited onto non-etched Si (100) wafers with low (a) and high (b) magnification.

Figure 2. XRD of iron silicon oxide nanowires deposited onto non-etched Si (100) wafer.

Figure 3. FT-IR spectrum of non-etched Si(100) wafer (a) and iron silicon oxide nanowires grown on it.

Figure 4. FESEM of iron silicon oxide nanowires deposited onto etched Si(100) wafer with low (a) and high (b) magnification.

Figure 5. XRD of iron silicon oxide nanowires deposited onto etched Si(100) wafer

Figure 6. FT-IR spectrum of etched Si(100) wafer (a) and iron silicon oxide nanowires grown on it.

Figure 7. FESEM of iron silicon oxide nanowires deposited onto etched Si(100) wafer with high magnification.

Figure 8. Proposed growth model based on the results presented in this paper.

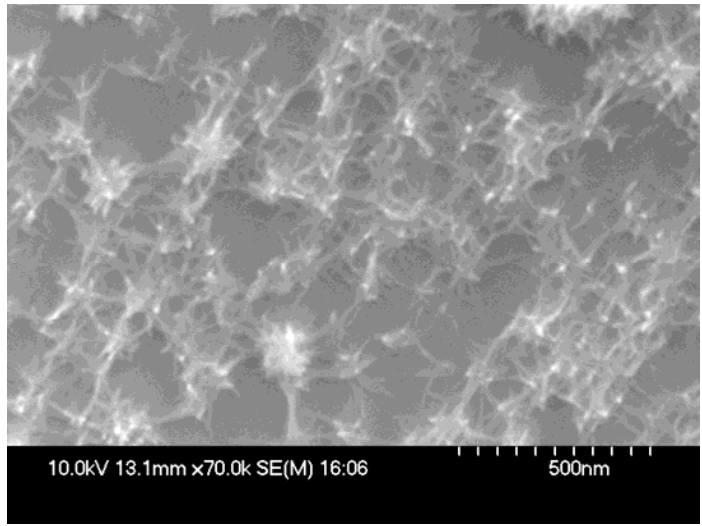

(a)

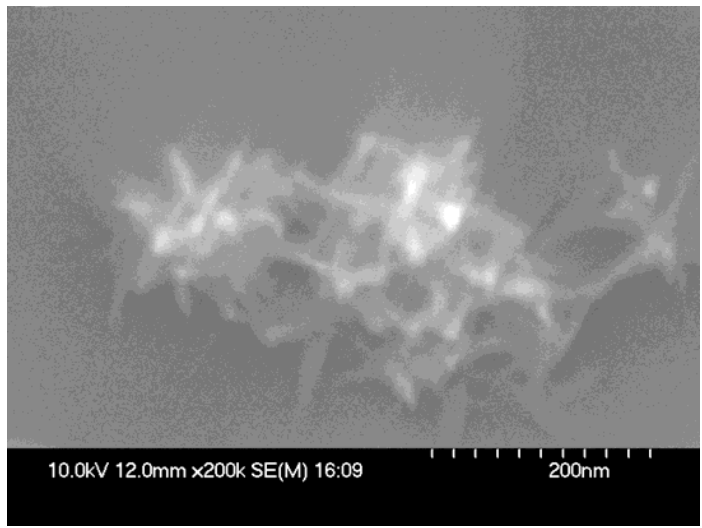

(b)

**Fig. 1** FESEM of iron silicon oxide nanowires deposited onto non-etched Si (100) wafer with low (a) and high (b) magnification.

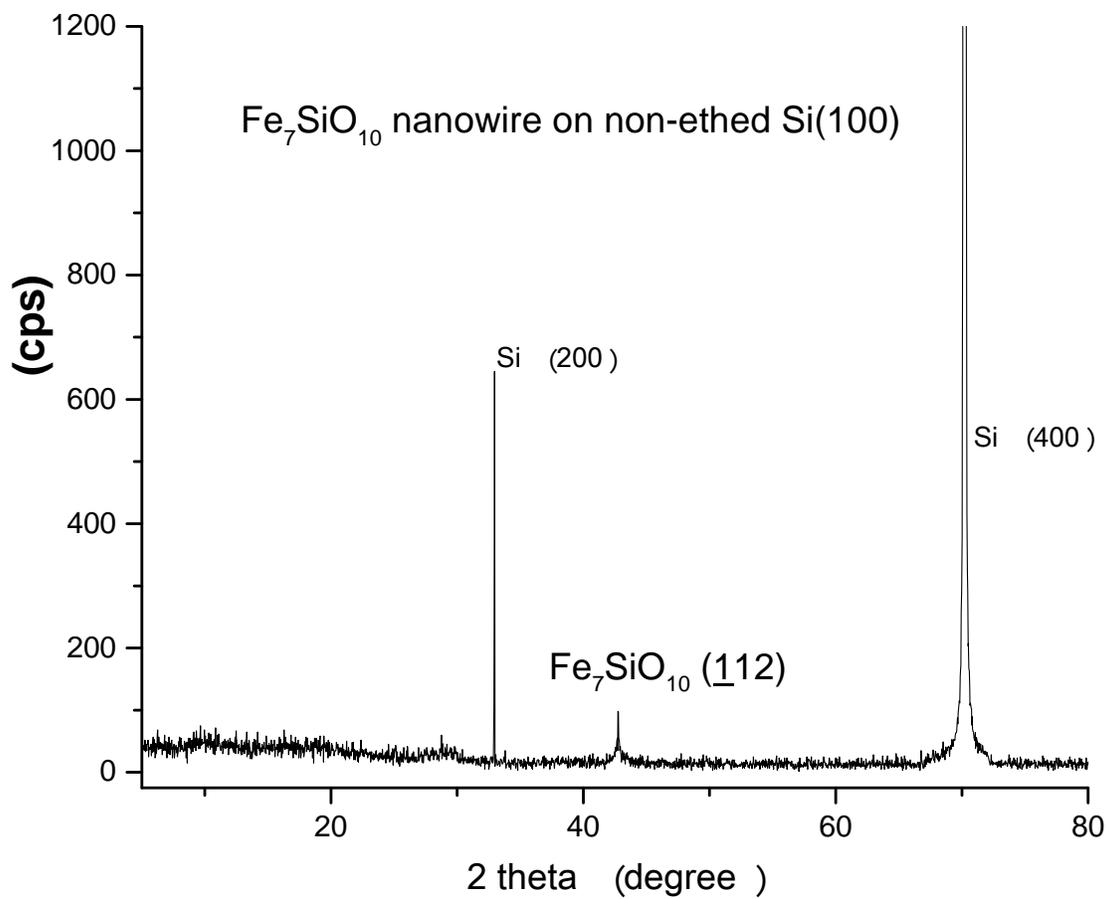

**Fig. 2** XRD of iron silicon oxide nanowires deposited onto non-etched Si(100) wafer.

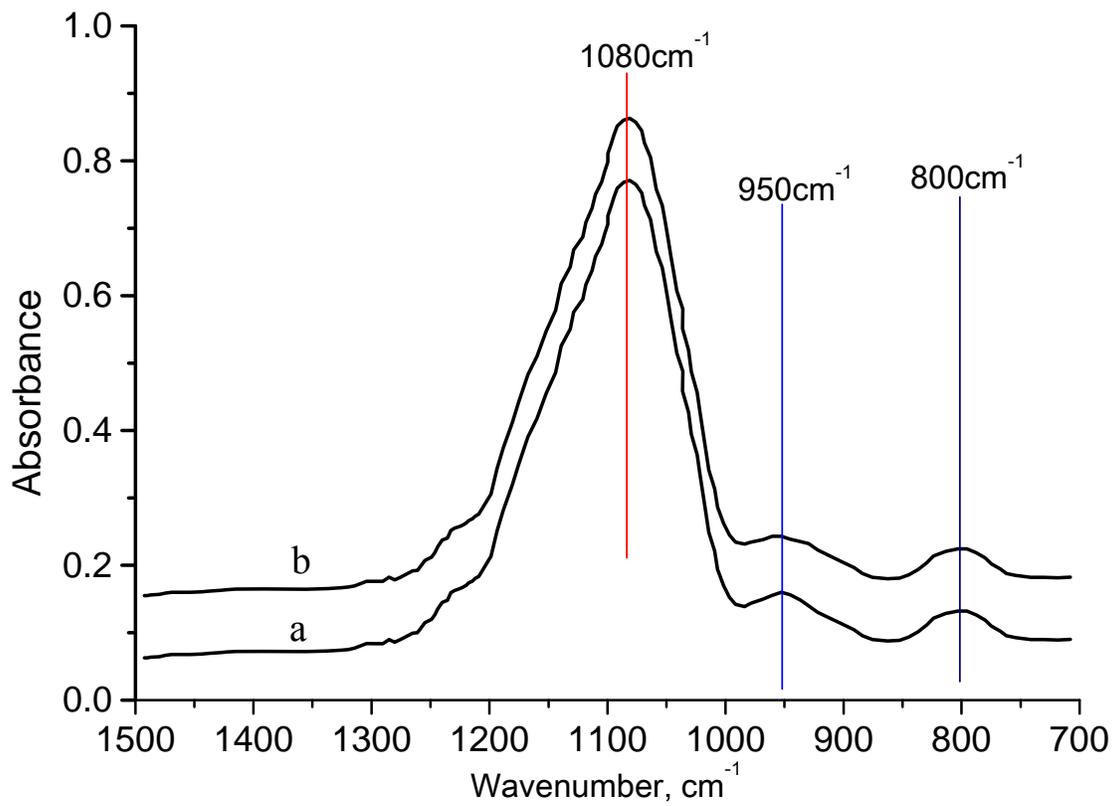

**Fig. 3** FT-IR spectrum of non-etched Si(100) wafer (a) and iron silicon oxide nanowires gown on it.

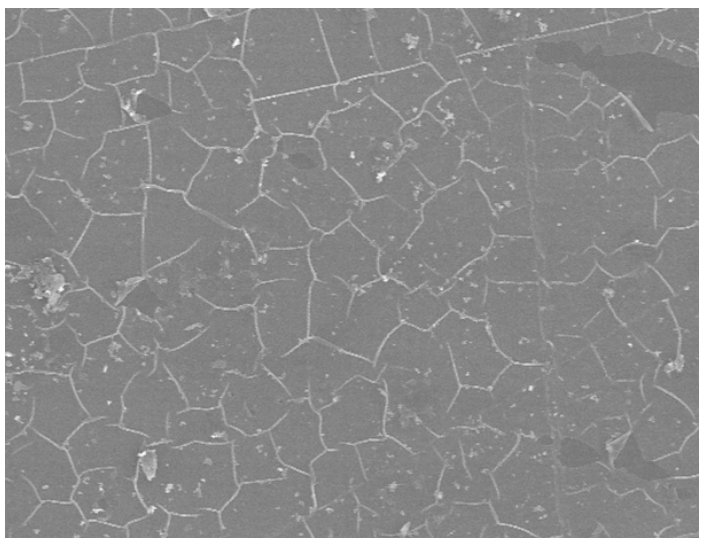

(a)

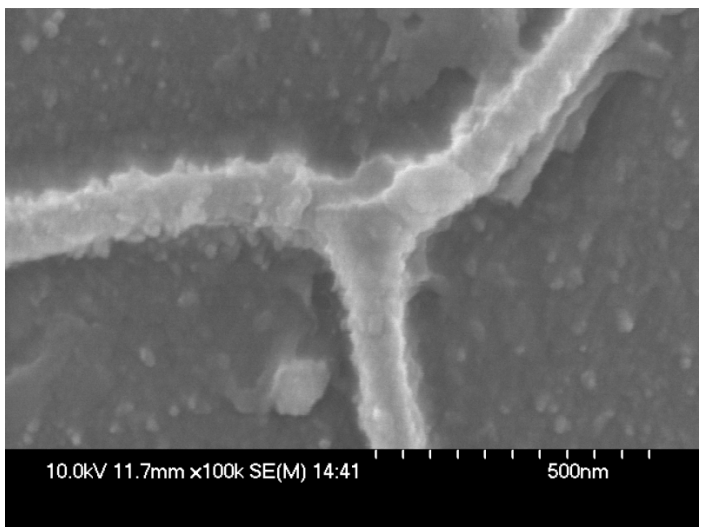

(b)

**Fig. 4** FESEM of iron silicon oxide nanowires deposited onto etched Si(100) wafer with low (a) and high (b) magnification.

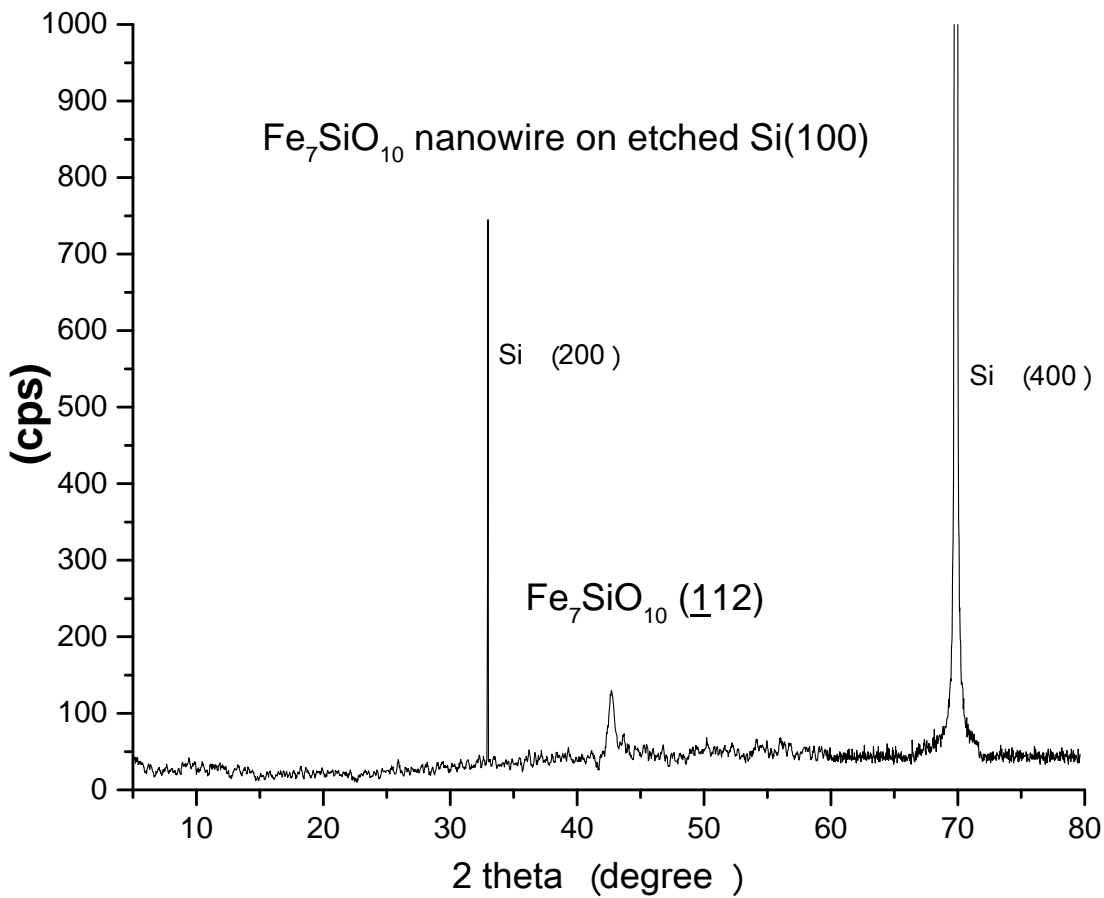

**Fig. 5** XRD of iron silicon oxide nanowires deposited onto etched Si(100) wafer.

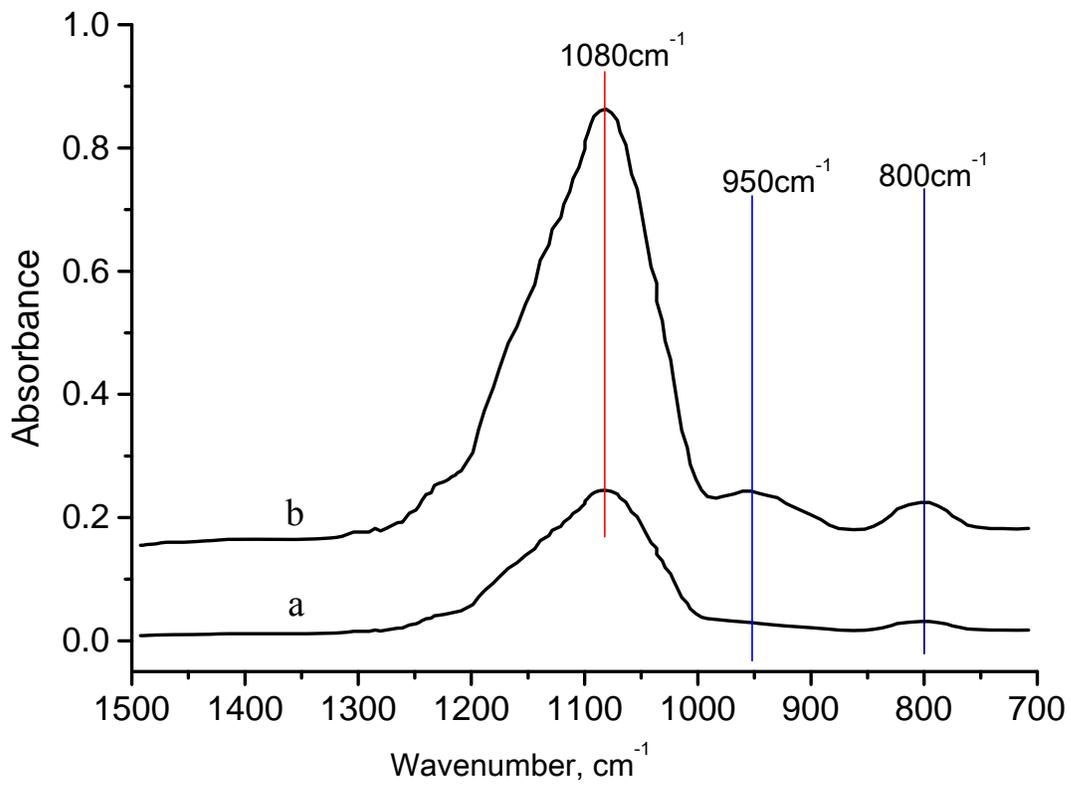

**Fig. 6** FT-IR spectrum of etched Si(100) wafer (a) and iron silicon oxide nanowires grown on it.

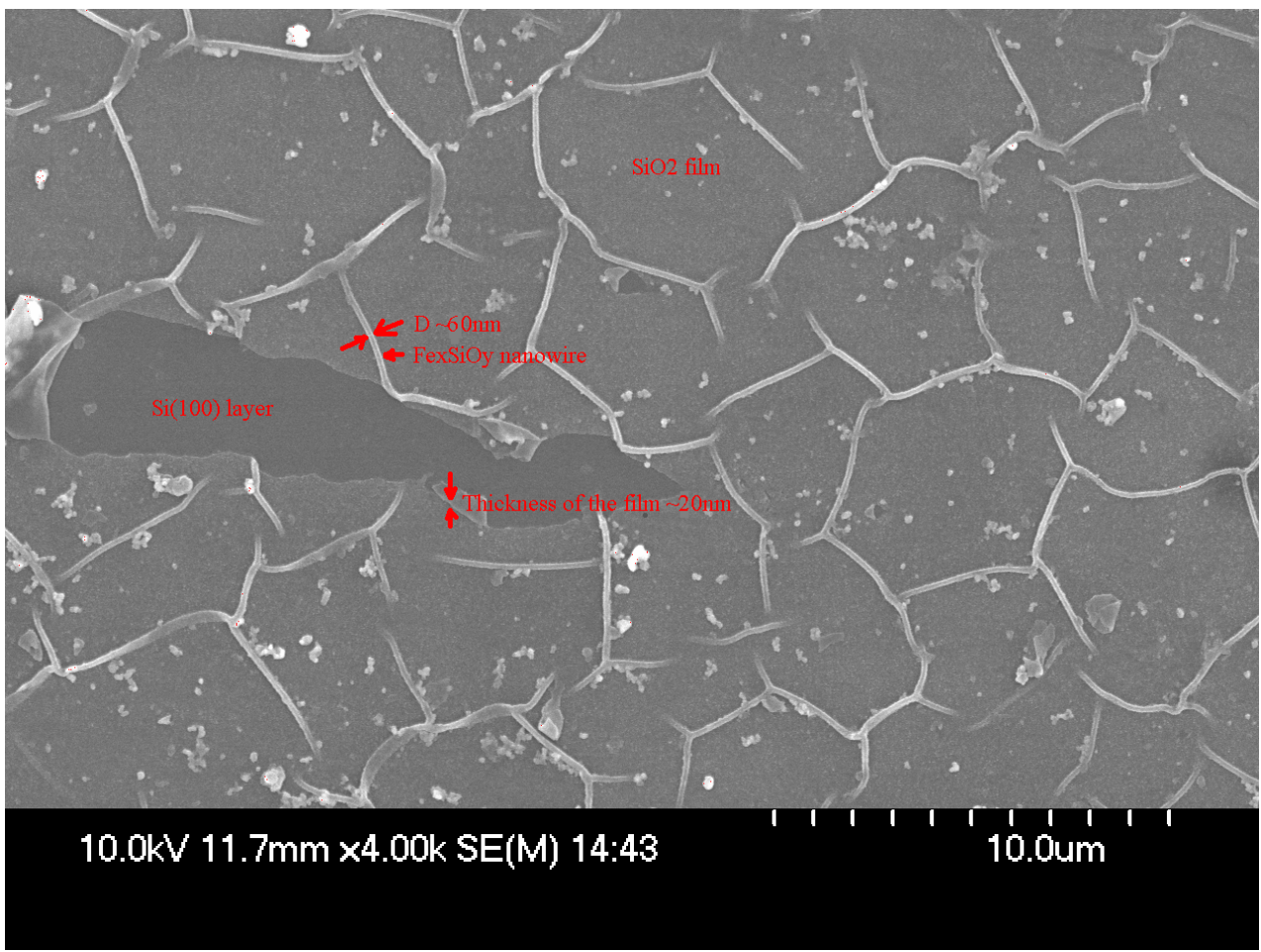

**Fig. 7** FESEM of iron silicon oxide nanowires deposited onto etched Si(100) wafer with high magnification.

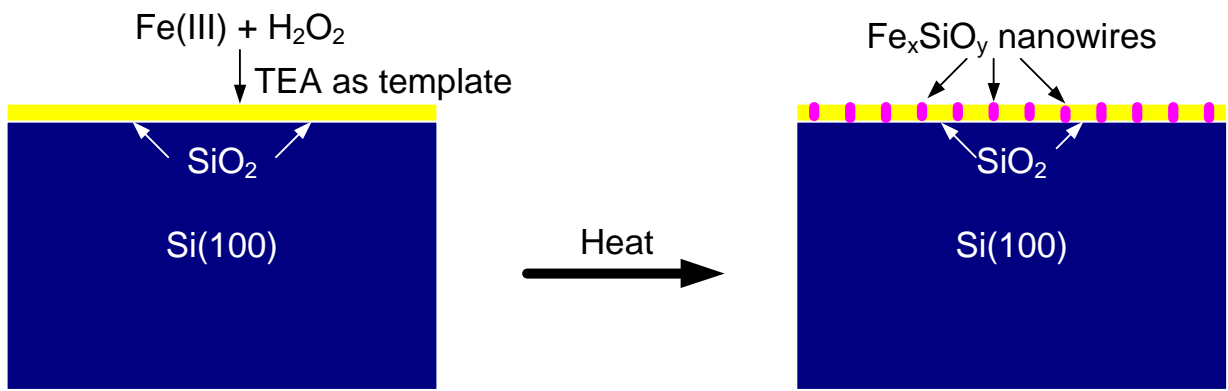

**Fig. 8** Proposed growth model based on the results presented in this paper.

**Supplementary Material (ESI) for Chemical Communications (2013)**

Fig. S1 presented the FE-SEM photographs of the as-prepared $Fe_7SiO_{10}$ nanowires self-assemble films deposited onto a non-etched Si(100) wafer. Please note Fig. S1(a), where there still existed some the $SiO_2$ layer with thickness of about ~50nm and the other parts there just left $Fe_7SiO_{10}$ nanowires film.

Fig. S2 presented the EDX spectrum of the film based on directly measuring the film on Si substrate and Tab. S1 showed the composition of the film. The carbon shown in Fig. S2 and Tab. S1 is from triethylamine which is either adsorbed in the $Fe_7SiO_{10}$ nanowires surface or residual inside the film.

Fig. S3 presented the FE-SEM photographs of the as-prepared $Fe_7SiO_{10}$ nanowires self-assemble films deposited onto etched Si(100) wafer.

Fig. S4 presented the EDX spectrum of the film based on directly measuring the film on Si substrate and Tab. S2 showed the composition of the film. The carbon is from triethylamine which is either adsorbed in the $Fe_7SiO_{10}$ nanowires surface or residual inside the film. Here the carbon containing is much higher than that of non-etched Si wafer situation, which may be explained as during the formation of the $SiO_2$ layer and much large amount of triethylamine was trapped inside it.

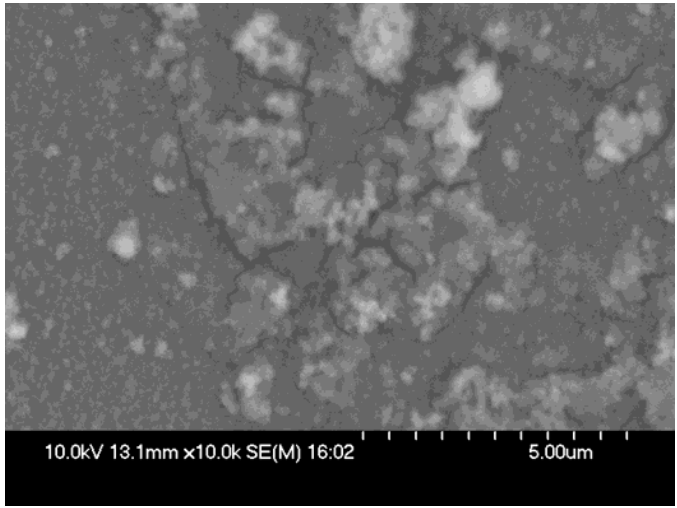 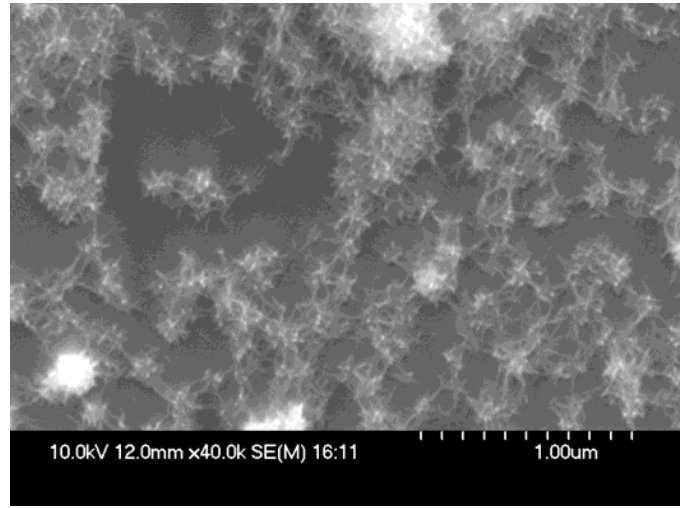

(a)   (b)

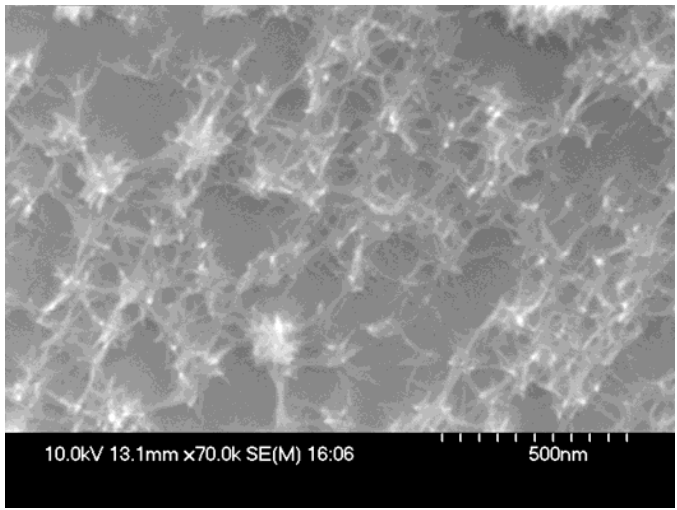 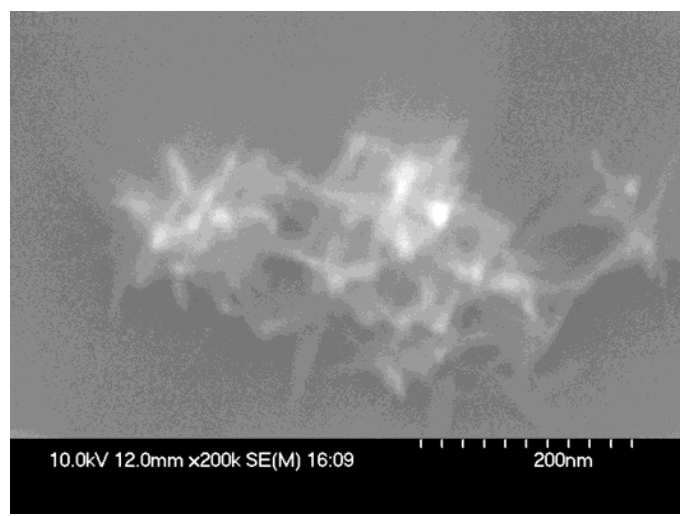

(c)   (d)

**Fig. S1** FESEM of iron silicon oxide nanowires onto non-etched Si(100) wafer from low to high magnification.

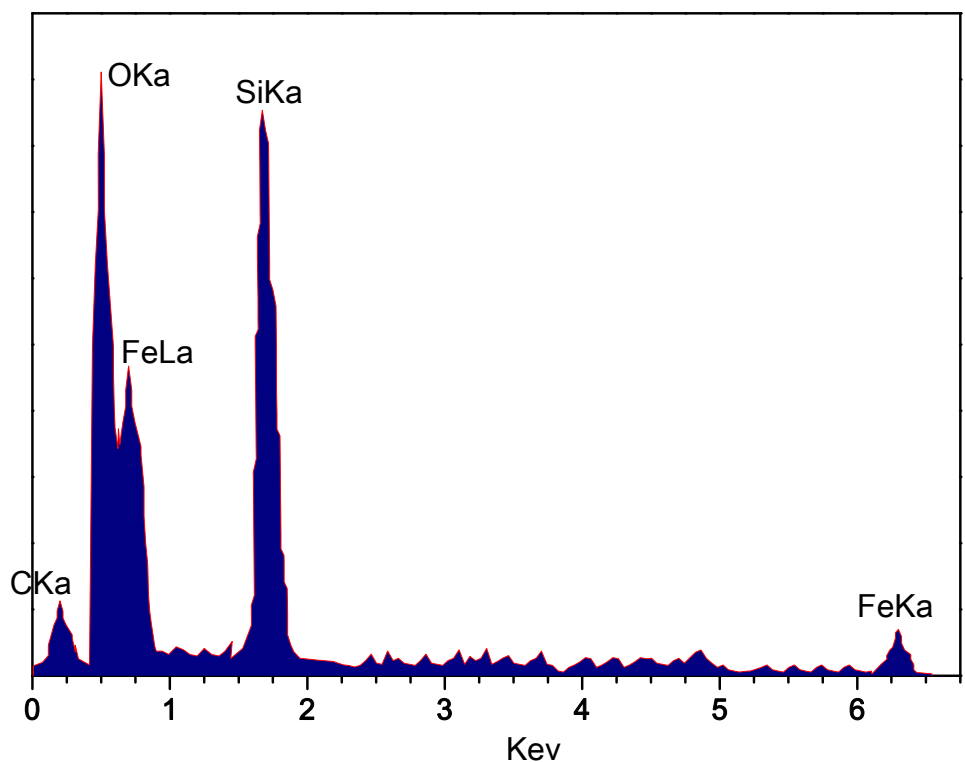

**Fig. S2** EDX of iron silicon oxide nanowires deposited on non-etched Si(100) wafer.

**Table S1** EDX result of iron silicon oxide nanowires onto non-etched Si(100) wafer

| Element | Wt % | At % |
| --- | --- | --- |
| C  K | 2.51 | 5.26 |
| O  K | 26.9 | 42.4 |
| Fe  L | 27.6 | 12.5 |
| Si  K | 42..99 | 39.84 |
| Total | 100.00 | 100.00 |

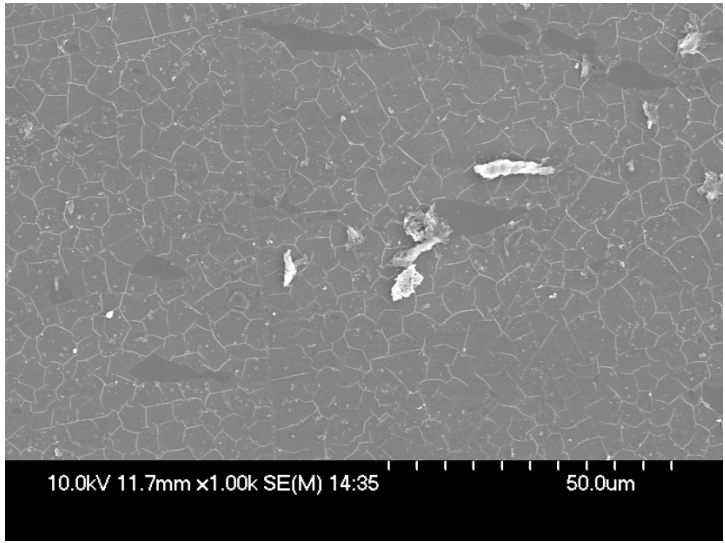

(a)

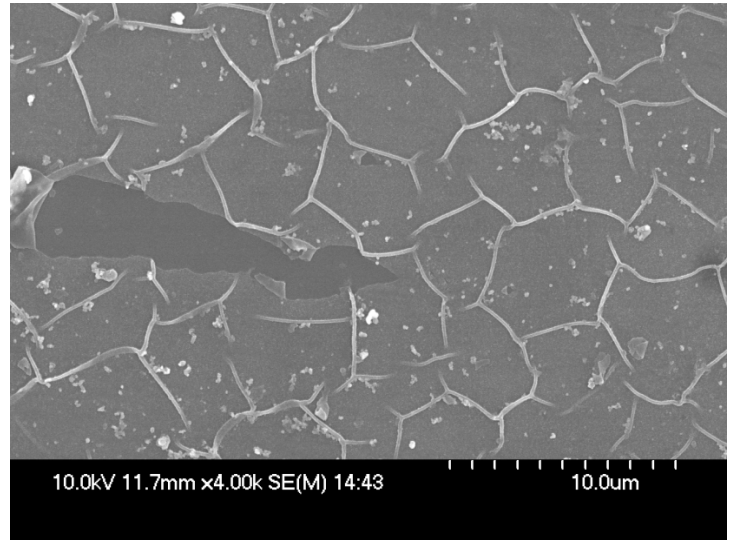

(b)

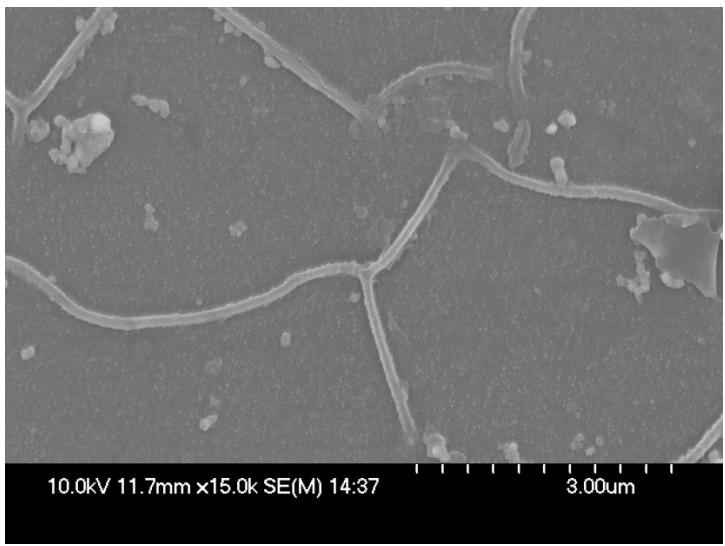

(c)

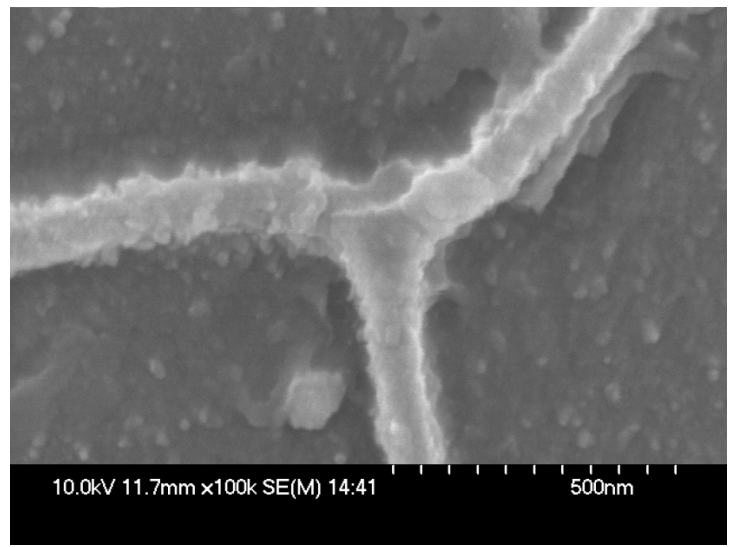

(d)

**Fig. S3** FESEM of iron silicon oxide nanowires onto etched Si(100) wafer from low to high magnification.

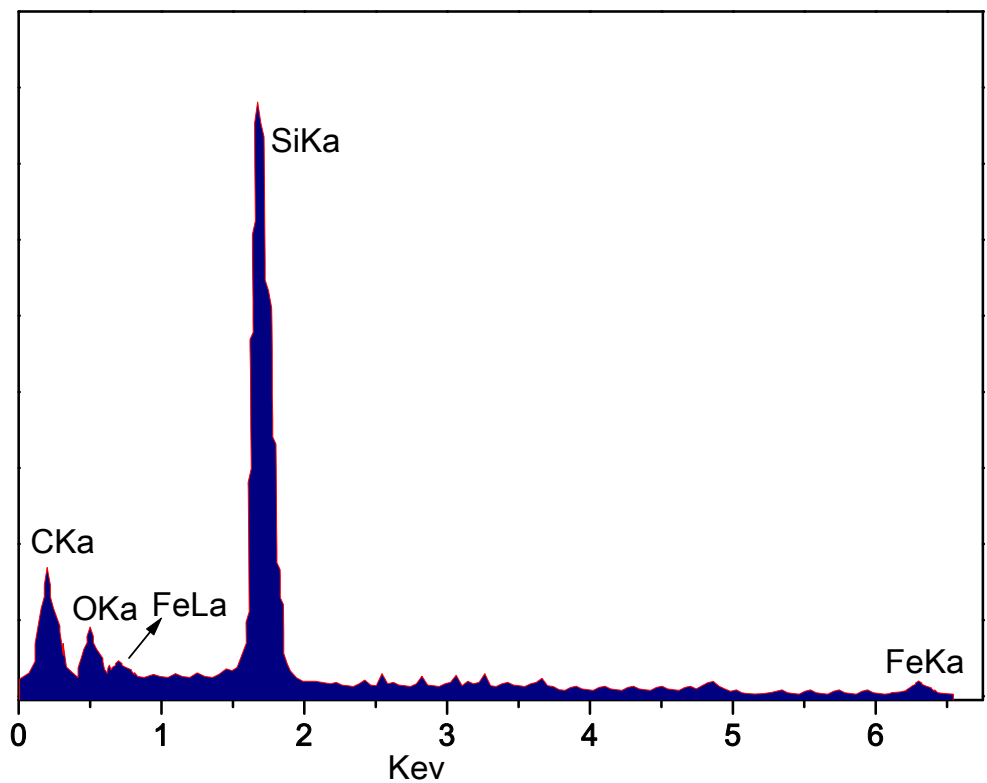

**Fig. S4** EDX of iron silicon oxide nanowires deposited onto etched Si(100) wafer.

**Table S2** EDX result of iron silicon oxide nanowires onto etched Si(100) wafer

| Element | Wt % | At % |
|---|---|---|
| C  K | 7.51 | 15.76 |
| O  K | 5.38 | 8.48 |
| Fe  L | 5.52 | 2.49 |
| Si  K | 81.59 | 73.26 |
| Total | 100.00 | 100.00 |